\documentclass[dvips,onecolumn,reprint,a4paper,superscriptaddress,prb]{revtex4}
\usepackage{amsmath}    
\usepackage{verbatim}   
\usepackage{color}      
\usepackage{subfigure}  
\usepackage{hyperref}   
\usepackage{bbold}
\usepackage{mathrsfs}
\usepackage{textcomp}

\begin{document}
\newcommand{\tat}{\textasciitilde{}}
\newcommand{\mk}{\mathbf{k}}

\title{Effective field theory description of topological crystalline insulators}
\author{N.M. Vildanov}
\affiliation{I.E.Tamm Department of Theoretical Physics,
P.N.Lebedev Physics Institute, 119991 Moscow, Russia}


\begin{abstract}
    We propose a phenomenological theory for topological
    crystalline insulators with time reversal and $C_4$
    symmetries. First, we introduce a fictitious space and
    transformation of electromagnetic field operators. This
    transformation leaves the speed of light unchanged but changes
    the elementary charge to $\sqrt{2}e$. Then we formulate the theory of
    topological crystalline insulators in terms of transformed fields in this fictitious
    space as 3D BF theory containing $\pi$-flux excitations. It is known that a 3D BF theory with half flux quantum
    excitations describes low energy properties of time reversal invariant insulators. By
    making an inverse transform we recover the effective field
    theory in original space. It turns out that this field theory
    contains quarter flux quantum excitations.
\end{abstract}

\maketitle

\section{Introduction}

In a recent paper \cite{fu}, Fu showed that there are 3D
topological band insulators characterized by a $Z_2$ index
protected by time reversal and $C_4$ symmetries. These insulators
have gapless surface states, 2D massless fermions with quadratic
spectrum, only on one of their surfaces (the one which is
perpendicular to the $C_4$ symmetry axis). Other surfaces of these
insulators turn out to be gapped.

Three-dimensional time reversal invariant insulators\cite{fkm,
moore, roy} are characterized by the following topological
response, \cite{tft,emv} the axion term $(c=1)$
\begin{equation}\label{axion}
    \frac{\theta e^2}{2\pi h}\mathbf{E}\cdot\mathbf{B}
\end{equation}
with $\theta=(2n-1)\pi$ in the non-trivial phase. Since the axion
term \eqref{axion} is related to a surface quantum Hall effect
with the conductivity $\frac{\theta}{2\pi}\frac{e^2}{h}$, and
since the Hall conductivity of massless 2D fermions with quadratic
spectrum is twice as bigger than that of massless 2D fermions with
linear spectrum, one may expect that electromagnetic response of
topological crystalline insulators of Ref.\onlinecite{fu} is given
by an axion term with $\theta=2\pi$. We show that this is indeed
the case by employing the technique for computation of the axion
term described in the paper \onlinecite{ryu}.

There is a topological field theory description of 3D time
reversal invariant insulators\cite{cho} which is given by 3D BF
theory
$$
\mathcal{L}_{BF}=\frac{k}{4\pi}\varepsilon^{\mu\nu\rho\lambda}a_\mu
\partial_\nu b_{\rho\lambda}+\frac{1}{2\pi}\varepsilon^{\mu\nu\rho\lambda}A_\mu
\partial_\nu b_{\rho\lambda}
$$
where $k=\frac{2}{2n-1}$. This theory can explain all of the main
properties of 3D time-reversal invariant insulators: $Z_2$-ness,
axion term, gapless edge theory. One may wonder is there an
analogous theory for crystalline insulators protected by point
group symmetries.\cite{fu}

In constructing such a theory one is faced with some difficulties
which are clear already at this stage of discussion. One of them
is related to the fact that $C_4$ symmetry of the underlying
microscopic model can not enter the topological field theory in an
explicit way and should be hidden. Second is related to the fact
that an axion term with $\theta=2\pi$ is equivalent to an axion
term with $\theta=0$ due to $2\pi$ periodicity,
\cite{wilczek,tft,emv,vazifeh} which makes it difficult to explain
$Z_2$-ness of topological crystalline insulators in the context of
topological field theory.

In this paper, we construct a theory that allows to circumvent
difficulties mentioned above. This theory explains $Z_2$
classification of topological crystalline insulators not in terms
of original fields, but in terms of fields transformed to a
fictitious space. It turns out that unit charge in this space is
different from that of an electron charge and as a consequence it
follows that $2\pi$ axion electrodynamics transforms to $\pi$
axion electrodynamics. This is crudely how $Z_2$ classification is
explained. This theory also contains traces of $C_4$ symmetry
because it supports quarter flux quantum excitations.

\section{Computation of the topological response}\label{theta}

Consider the following hamiltonian
$$
H(k_x,k_y,k_z)=\hat{h}(k_x,k_y)\otimes\tau_z+M\cdot
\mathbb{1}\otimes\tau_x-k_z\cdot \mathbb{1}\otimes\tau_y
$$
where
\begin{equation}\label{h}
    \hat{h}(k_x,k_y)=\frac{k^2}{2m} +\frac{1}{2m_1}(k_x^2-k_y^2)\cdot
\hat{\sigma}_z+\frac{1}{m_2}k_x k_y \cdot \hat{\sigma}_x
\end{equation}
If $M>0$ for $z>0$ then (in the $\tau$ space) z-dependence of the
wavefunction is
$$
\psi=\begin{pmatrix}
  0 \\
  1 \\
\end{pmatrix}e^{-Mz}
$$
Analogously, if $M<0$ for $z<0$ then
$$
\psi=\begin{pmatrix}
  0 \\
  1 \\
\end{pmatrix}e^{|M|z}
$$
This means that there is a 2D massless chiral fermion with
quadratic spectrum \eqref{h} localized near the surface which is
perpendicular to the symmetry axis and which separates two models
with different sign of mass $M$.

There are two symmetries of the hamiltonian, time reversal
operation $T$ which coincides with complex conjugation $K$, and
$C_4$ rotation in the $(k_x,k_y)$ plane $U=\hat{\sigma}_y\otimes
1$, $(UT)^2=-1$:
$$
T^{-1}H(k_x,k_y,k_z)T=H(-k_x,-k_y,-k_z),\qquad
U^{-1}H(k_x,k_y,k_z)U=H(k_y,-k_x,k_z)
$$

We will compute the axion term assuming that
$\hat{h}(k_x,k_y)=\frac{1}{2m_1}(k_x^2-k_y^2)\cdot
\hat{\sigma}_z+\frac{1}{m_2}k_x k_y \cdot \hat{\sigma}_x$, where
$m_1=m_2$, using the approach given in the appendix of Ref.
\onlinecite{ryu}. Calculation can be easily generalized for
unequal masses $m_1\neq m_2$, but unfortunately we could not do
the calculation for generic $h$, Eq. \eqref{h}. Lagrangian in the
euclidean space-time is
$$
    \overline{\psi}(\gamma_\mu d_\mu-M)\psi
$$
where
$$
    \gamma_0=\mathbb{1}\otimes \hat{\tau}_x, \quad
    \gamma_1=\sigma_z\otimes{\tau}_y, \quad
    \gamma_2=\sigma_x\otimes{\tau}_y,\quad
    \gamma_3=\mathbf{1}\otimes{\tau}_z,\quad
    \gamma_5=\sigma_y\otimes{\tau}_y=\gamma_0\gamma_1\gamma_2\gamma_3
$$
$$
    d_0=\partial_t+iA_0,\quad d_1=\frac{i}{2m}(k_y^2-k_x^2), \quad
    d_2=-\frac{i}{2m}(k_xk_y+k_yk_x),\quad d_3=\partial_z+iA_z,
    \quad k_{x,y}=\partial_{x,y}+iA_{x,y}
$$

$M\rightarrow M e^{i\gamma_5}$ interpolates between hamiltonians
with different sign of mass when $\alpha$ varies from $0$ to
$\pi$, violating time reversal symmetry for intermediate values of
$\alpha$. If $\alpha$ slowly varies in space such that there is a
domain wall on which the mass $M$ flips the sign, then the action
on one of the sides of the domain wall will contain an additional
term with respect to the action on the opposite side. This
additional term is an anomaly due to the change of the Jacobian
under the transformation $\psi^\prime=e^{i\alpha\gamma_5/2}\psi$.
This anomaly can be computed by a proper regularization, using
Fujikawa method, and is given by
\begin{equation}\label{anomaly}
    i\pi \int\frac{d^4k}{(2\pi)^4}~{\rm{tr}}\{\gamma_5 f[\langle k|(\gamma_\mu
    d_\mu)^2/M^2|k\rangle]\}
\end{equation}
where the function $f(s)$ satisfies the requirements
$$
    f(0)=1,\quad f(\infty)=0,\quad sf^\prime(s)=0\quad {\rm{when}}\quad s=0,\infty
$$
It is easy to show that $(\gamma_\mu d_\mu)^2=d_\mu
d_\mu+1/4[\gamma_\mu,\gamma_\nu][d_\mu,d_\nu]$. Since
${\rm{tr}}\{\gamma_5[\gamma_\mu,\gamma_\nu][\gamma_\rho,\gamma_\sigma]\}=16\varepsilon_{\mu\nu\rho\sigma}$,
in the absence of the gauge fields, $A_\mu=0$, one obtains
$\langle k|-(d_\mu
d_\mu)|k\rangle=k_0^2+k_z^2+(k_x^2+k_y^2)^2/4m^2$. To obtain a
non-zero result in \eqref{anomaly} in the limit
$M\rightarrow\infty$ we need to expand $f$ up to the second order
in $[\gamma_\mu,\gamma_\nu][d_\mu,d_\nu]/M^2$ and retain the terms
in $[d_\mu,d_\nu][d_\rho,d_\sigma]$ which contain the operators
$\partial^2_x$ and $\partial^2_y$. Namely, the terms that should
be retained are
\begin{equation}\label{dd1}
    [d_0,d_3]=iE_z,\quad
    [d_1,d_2]=\frac{i}{m^2}B_z(\partial^2_x+\partial^2_y)+...
\end{equation}
\begin{equation}\label{dd2}
    [d_0,d_1]=\frac{1}{m}(E_y\partial_y-E_x\partial_x)+...,\quad
[d_3,d_2]=\frac{1}{m}(B_y\partial_y-B_x\partial_x)+...
\end{equation}
\begin{equation}\label{dd3}
    [d_0,d_2]=\frac{1}{m}(E_x\partial_y+E_y\partial_x)+...,\quad
[d_3,d_1]=-\frac{1}{m}(B_y\partial_x+B_x\partial_y)+...
\end{equation}
Then \eqref{anomaly} becomes
$$
    i\pi \int\frac{d^4k}{(2\pi)^4}\cdot 4\cdot\frac{1}{M^4} \langle
    k|\left\{[d_0,d_1][d_2,d_3]-[d_0,d_2][d_1,d_3]+[d_0,d_3][d_1,d_2]\right\}|k\rangle f^{\prime\prime}(s)=i\pi \int\frac{d^4k}{(2\pi)^4}\cdot
    4\cdot\frac{1}{M^4m^2}\cdot (\mathbf{B}\mathbf{E})\cdot
    (k_x^2+k_y^2)f^{\prime\prime}(s)
$$
where $s=k_0^2+k_z^2+(k_x^2+k_y^2)^2/4m^2$. Calculating the
integral, which does not depend on the specific choice of the
function $f$, we get the axion term $i\frac{\theta}{2\pi
h}(\mathbf{B}\mathbf{E})$ with $\theta=2\pi$.

\section{Transformation of electromagnetic field}\label{tilde_electrodynamics}

In this section we introduce a fictitious space (denoted in the
following by \textasciitilde{}) and a transformation of
electromagnetic field which is motivated by the previous
discussion, Eqs.(\ref{dd1}--\ref{dd3}). The aim is to reformulate
the theory in this fictitious space and in terms of transformed
electromagnetic field operators.

It is more convenient to work in momentum space. Momentum
operators in \textasciitilde{} space are given by
\begin{equation}\label{momentum}
    K_x=\frac{k_x^2-k_y^2}{k_\perp},\qquad
K_y=\frac{2k_xk_y}{k_\perp},\qquad K_z=k_z \qquad (k_y>0)
\end{equation}
and with reverse sign for $k_y<0$. From this definition one can
see that the absolute value of the momentum does not change since
$K_x^2+K_y^2=k_\perp^2$, and that the vector $\mathbf{K}$ covers
\textasciitilde{} momentum space twice. This means that
electromagnetic field operators in \textasciitilde{} space split
into two fields (generally, every field that can be transformed
into \tat{} space will split into two fields). We will divide
$\mathbf{k}_\perp$ plane into two regions: 1)
${\rm{sgn}}k_x={\rm{sgn}}k_y$ $(l=1)$ and 2)
${\rm{sgn}}k_x=-{\rm{sgn}}k_y$ $(l=2)$. The relation between
$\mathbf{k}$ and $\mathbf{K}$ integrations is given by
\begin{equation}\label{k_integral}
    2\int d^3k=\sum_{l=1,2} \int d^3K
\end{equation}

Scalar field in \tat{} space is defined according to relations
\begin{equation}\label{tilde_scalar}
    \tilde{\phi}^{(1)}(\mathbf{K})=\frac{\phi(\mathbf{k})}{\sqrt{2}},\quad
    (k_x,k_y>0);\qquad \tilde{\phi}^{(2)}(\mathbf{K})=\frac{\phi(\mathbf{k})}{\sqrt{2}},\quad
    (-k_x,k_y>0)
\end{equation}
and $\tilde{\phi}^{(l)}(\mathbf{-K})$ is defined as complex
conjugate of Eqs.\eqref{tilde_scalar}, such that in accordance
with \eqref{k_integral} we have
$$
    \int \phi(\mathbf{k})\phi(\mathbf{-k}) d^3k=\sum_{l=1,2} \int \tilde{\phi}^{(l)}(\mathbf{K})\tilde{\phi}^{(l)}(\mathbf{-K}) d^3K
$$
Transformation formulas for vector potential must be consistent
with Eq. \eqref{tilde_scalar}. They are defined as follows
\begin{equation}\label{tilde_vector}
    \tilde{{A}}^{(l)}_{x}(\mathbf{K})=\frac{k_x{A}_x(\mathbf{k})-k_y{A}_y(\mathbf{k})}{\sqrt{2}k_\perp},\qquad
    \tilde{{A}}^{(l)}_{y}(\mathbf{K})=\frac{k_y{A}_x(\mathbf{k})+k_x{A}_y(\mathbf{k})}{\sqrt{2}k_\perp},\qquad
    \tilde{{A}}^{(l)}_{z}(\mathbf{K})=\frac{{A}_z(\mathbf{k})}{\sqrt{2}}
\end{equation}
and $\tilde{\mathbf{A}}^{(l)}(-\mathbf{K})$ as complex conjugate
of \eqref{tilde_vector}. One can see from Eq. \eqref{tilde_vector}
that when vector potential in original space is a pure gauge, then
it is pure gauge also in \tat{} space, which is quite reasonable.
Transformation formulas for electric and magnetic fields can be
obtained by substituting \eqref{tilde_scalar} and
\eqref{tilde_vector} into equations $\tilde{E}_i=-\partial_t
\tilde{A}_i-\tilde{\partial}_i\tilde{\phi}~$,$\tilde{H}_i=\varepsilon_{ijk}\tilde{\partial}_j\tilde{A}_k$.
The result is that $\tilde{\mathbf{E}}$ and ${\mathbf{E}}$,
$\tilde{\mathbf{H}}$ and ${\mathbf{H}}$ are related to each other
exactly by the same formulas as $\tilde{\mathbf{A}}$ and
${\mathbf{A}}$, Eq. \eqref{tilde_vector}, and
$\tilde{\mathbf{E}}^{(l)}(-\mathbf{K})$,
$\tilde{\mathbf{H}}^{(l)}(-\mathbf{K})$ by their complex
conjugates:
\begin{equation}\label{tilde_e}
    \tilde{{E}}^{(l)}_{x}(\mathbf{K})=\frac{k_x{E}_x(\mathbf{k})-k_y{E}_y(\mathbf{k})}{\sqrt{2}k_\perp},\qquad
    \tilde{{E}}^{(l)}_{y}(\mathbf{K})=\frac{k_y{E}_x(\mathbf{k})+k_x{E}_y(\mathbf{k})}{\sqrt{2}k_\perp},\qquad
    \tilde{{E}}^{(l)}_{z}(\mathbf{K})=\frac{{E}_z(\mathbf{k})}{\sqrt{2}}
\end{equation}
\begin{equation}\label{tilde_h}
    \tilde{{H}}^{(l)}_{x}(\mathbf{K})=\frac{k_x{H}_x(\mathbf{k})-k_y{H}_y(\mathbf{k})}{\sqrt{2}k_\perp},\qquad
    \tilde{{H}}^{(l)}_{y}(\mathbf{K})=\frac{k_y{H}_x(\mathbf{k})+k_x{H}_y(\mathbf{k})}{\sqrt{2}k_\perp},\qquad
    \tilde{{H}}^{(l)}_{z}(\mathbf{K})=\frac{{H}_z(\mathbf{k})}{\sqrt{2}}
\end{equation}
It is also possible to incorporate source terms (current $j^\mu$)
into the theory, transformation properties of which are analogous
to \eqref{tilde_scalar},\eqref{tilde_vector}.

One of the consequences of Eqs. \eqref{tilde_e} and
\eqref{tilde_h} is that scalar products in two spaces differ by a
factor of 2. For instance,
$\tilde{\mathbf{E}}(\mathbf{K})\tilde{\mathbf{E}}(\mathbf{-K})=\mathbf{E}(\mathbf{k})\mathbf{E}(\mathbf{-k})/2$,
and analogous formulas are true also for the scalar product of the
pairs $\tilde{\mathbf{H}}(\mathbf{K})$ and
$\tilde{\mathbf{H}}(\mathbf{-K})$,
$\tilde{\mathbf{E}}(\mathbf{K})$ and
$\tilde{\mathbf{H}}(\mathbf{-K})$. This leads to equalities
$$
    \int \mathbf{E}^2 d^3x=\sum_{l=1,2}\int
    [\tilde{\mathbf{E}}^{(l)}]^2d^3x,\qquad \int \mathbf{H}^2 d^3x=\sum_{l=1,2}\int
    [\tilde{\mathbf{H}}^{(l)}]^2d^3x
$$
which mean, in particular, that lagrangian density of
electromagnetic field does not change its form and, therefore,
$\tilde{A}$ is a true electromagnetic potential in \tat{} space.
Electric and magnetic fields \eqref{tilde_e} and \eqref{tilde_h}
defined in this way satisfy Maxwell's equations with the same
speed of light $c$.

Coulomb gauge condition $\mathbf{k}\mathbf{A}(\mathbf{k})=0$
transforms to $\mathbf{K}\tilde{\mathbf{A}}(\mathbf{K})=0$. So, it
can be checked by direct computation that the commutation
relations\cite{qed}
$[{E_i}(\mathbf{k}),{A_j}(\mathbf{-k}^\prime)]=4\pi i
\left(\delta_{ij}-\frac{k_ik_j}{k^2}\right)\cdot
(2\pi)^3\delta(\mathbf{k}-\mathbf{k}^\prime)$, which are satisfied
in quantum theory in the Coulomb gauge, are transformed to
$[{\tilde{E}_i^{(l)}}(\mathbf{K}),{\tilde{A}_j^{(l^\prime)}}(\mathbf{-K}^\prime)]=4\pi
i \left(\delta_{ij}-\frac{K_iK_j}{k^2}\right)\cdot
(2\pi)^3\delta(\mathbf{K}-\mathbf{K}^\prime)\delta_{ll^\prime}$.
This means that transformation formulas \eqref{tilde_vector} are
correctly defined also in quantum theory.

To see how the axion term transforms one needs to consider
$\theta$ varying in space. If $\theta=\theta(z)$ then, rewriting
the integral in momentum space, it can be shown that
$$
    \int\theta(z)\mathbf{E}\mathbf{H}~d^3x=\sum_{l}\int\theta(z)\tilde{\mathbf{E}}^{(l)}\tilde{\mathbf{H}}^{(l)}~d^3x
$$
So one can write
\begin{equation}\label{axion_transformation}
    \int_M
    \mathbf{E}\mathbf{H}=\sum_{l}\int_{\tilde{M}}\tilde{\mathbf{E}}^{(l)}\tilde{\mathbf{H}}^{(l)}
\end{equation}

\section{Some other properties of the fictitious electrodynamics}

One of the consequences of Eq. \eqref{tilde_scalar} is that the
scalar potential of a static point charge $q$,
$\phi({\mathbf{k}})=4\pi q/k^2$, becomes
$\tilde{\phi}({\mathbf{K}})=4\pi \tilde{q}/K^2$,
$\tilde{q}=q/\sqrt{2}$, i.e. scalar potential of a static point
charge $q/\sqrt{2}$. More generally, a simple correspondence can
be made between electric field distributions in the two spaces
when the field configuration has axial symmetry. However, this
does not mean that elementary charge in \tat{} space $\tilde{e}$
is given by $1/\sqrt{2}$ units of electronic charge.

In the same way, one can obtain the transformation $q_m\to
\tilde{q}_m=q_m/\sqrt{2}$, $\phi\to \tilde{\phi}=\phi/\sqrt{2}$
for a magnetic pole $q_m$ and a magnetic flux vertex $\phi$
parallel to $z$ axis.

There is one more configuration which has simple transformation
properties, a skyrmion. Consider the simplest skyrmion in 2+1D
\begin{equation}\label{skyrmion_original}
    \phi_0(r,\varphi)=\cos f(r),\quad \phi_x(r,\varphi)=\sin
    f(r)\cos\varphi,\quad \phi_y(r,\varphi)=\sin
    f(r)\sin\varphi
\end{equation}
This can be written in more convenient form as
\begin{equation*}
    \phi_0=\cos f(r),\quad \phi_x=\partial_x
    \int_0^{r} \sin f(r^\prime)dr^\prime,\quad \phi_y=\partial_y
    \int_0^{r} \sin f(r^\prime)dr^\prime.
\end{equation*}
Transformation formulas for $\phi_i$ are analogous to
\eqref{tilde_vector} (we drop the factor $\frac{1}{\sqrt{2}}$
because it does not affect the final result), and in coordinate
space we obtain
\begin{equation}\label{skyrmion_tilde}
    \tilde{\phi}_0^{(l)}(r,\varphi)=\cos f(r),\quad \tilde{\phi}_x^{(l)}(r,\varphi)=\sin
    f(r)\cos\varphi,\quad \tilde{\phi}_y^{(l)}(r,\varphi)=\sin
    f(r)\sin\varphi,\qquad l=1,2,
\end{equation}
where $r$ and $\varphi$ are polar coordinates in \tat{} space.

When a skyrmion is coupled to a doublet of two-component fermions
through the interaction lagrangian
$g\sigma_z(\boldsymbol{\phi}\hat{\tau})$, it acquires quantum
numbers, charge and spin.\cite{goldstone,chen,abanov} For the
skyrmion coupled to Dirac fermions induced fermion current is
given by the topological current of the skyrmion (we put number of
flavours $N_f=1$)
\begin{equation}\label{top_current}
    J_\mu=\frac{1}{8\pi}\varepsilon_{\mu\nu\lambda}\varepsilon_{ijk}n_i\partial_\nu
    n_j\partial_\lambda n_k
\end{equation}
where $n_i=\phi_i/|\boldsymbol{\phi}|$, and spin is determined by
the Hopf term
\begin{equation}\label{hopf}
    S_{Hopf}=i\pi\int d^3x J^\mu C_\mu
\end{equation}
where the gauge field $C_\mu$ is related to the current through
the relation $J_\mu=\varepsilon_{\mu\nu\lambda}\partial_\nu
C_\lambda$. Charge is given by the temporal component of
\eqref{top_current}
\begin{equation*}\label{charge}
    Q=\int d^2x \frac{1}{4\pi} \varepsilon_{ijk}n_i\partial_x
    n_j\partial_y n_k
\end{equation*}
The skyrmion configuration \eqref{skyrmion_original} has charge
$Q=e$ and spin $s=\frac{1}{2}$. We note that spin can not be
treated according to quasiclassical equations
\eqref{tilde_vector}. Correct quasiclassical treatment of the spin
of a skyrmion is as in Refs.\onlinecite{witten,zee,wu}.

However, we are interested in skyrmions in fermion models with
quadratic band touching. These were studied in\cite{moon} and the
main result can be formulated as the doubling of quantum numbers
compared to fermions with linear spectrum. Thus, we have that the
skyrmion \eqref{skyrmion_original} coupled to fermions with
quadratic spectrum has charge $Q=2e$ and spin $s=1$. This is
analogous to doubling of the $\theta$ term that we have found in
Sec. \ref{theta}. After transformation to \tat{} space, the Hopf
term $S_{Hopf}=2i\pi\int d^3x J^\mu C_\mu$ of such a skyrmion will
be equally distributed between skyrmions \eqref{skyrmion_tilde}:
$S_{Hopf}=\tilde{S}_{Hopf}^{(1)}+\tilde{S}_{Hopf}^{(2)}$,
$\tilde{S}_{Hopf}^{(l)}=i\pi\int d^3x \tilde{J}^\mu
\tilde{C}_\mu$, where
$\tilde{J}_\mu=\frac{1}{8\pi}\varepsilon_{\mu\nu\lambda}\varepsilon_{ijk}\tilde{n}_i\tilde{\partial}_\nu
    \tilde{n}_j\tilde{\partial}_\lambda \tilde{n}_k$, $\tilde{n}_i=\tilde{\phi}_i/|\tilde{\boldsymbol{\phi}}|$
(we note that fermionic charge current is
$\tilde{j}_\mu=\sqrt{2}e\tilde{J}_\mu$). This means that skyrmions
in \tat{} space will have half of the spin of the skyrmion in
original space. In the case under consideration, their spin is
$\frac{1}{2}$, i.e. they are fermions.

We saw that the correspondence
$S_{Hopf}=\tilde{S}_{Hopf}^{(1)}+\tilde{S}_{Hopf}^{(2)}$ holds for
the skyrmions \eqref{skyrmion_original} and
\eqref{skyrmion_tilde}, but it can be shown that it is more
general. One can see that $\tilde{\phi}_i$ depends continuously on
$\phi_i$. If we continuously deform $\phi_i$ then $S_{Hopf}$ does
not change, since it is topological invariant. Analogously,
$\tilde{S}_{Hopf}^{(l)}$ does not change under continuous
deformations of $\tilde{\phi}_i$. So
$S_{Hopf}=\tilde{S}_{Hopf}^{(1)}+\tilde{S}_{Hopf}^{(2)}$ remains
valid for continuous deformations of the skyrmions
\eqref{skyrmion_original} and \eqref{skyrmion_tilde}.

Thus, we have established that transformation of spin-1 bosonic
skyrmions with electric charge $2e$  to \tat{} space results in
two different types of spin-$\frac{1}{2}$ fermions each having
charge $\sqrt{2}e$. Suppose that we interpret this as
$\tilde{e}=\sqrt{2}e$. Since $\theta$ term is proportional to the
square of elementary charge, this means that the value of the
$\theta$-angle in \tat{} space is two times smaller than in
original space. This idea is pursued further in the next section
from different point of view.

Now we consider an example for illustration, an axially symmetric
charge distribution $\rho(r)$ rotating around the symmetry axis
$z$ with small angular velocity $\omega$ ($\omega a\ll c$, $a$ is
characteristic size in transverse direction). We have
$j_x=-\omega\partial_yg(r)$, $j_y=\omega \partial_xg(r)$, $j_z=0$,
where $g(r)=\int_0^r\rho(r^\prime)dr^\prime$. Transformation to
\tat{} space gives
$\tilde{\rho}^{(l)}(r)=\frac{1}{\sqrt{2}}\rho(r)$,
$\tilde{j}_x^{(l)}=-\omega\tilde{\partial}_y\tilde{g}(r)$,
$\tilde{j}_y^{(l)}=\omega\tilde{\partial}_x\tilde{g}(r)$, $l=1,2$
where $\tilde{g}(r)=\int_0^r\tilde{\rho}(r^\prime)dr^\prime$. This
means that angular velocity in \tat{} space is the same as in
original space. In particular, rotation around $z$ axis in
original space of an axially symmetric configuration results in
rotation by the same angle in \tat{} space. Wavefunction of a
bosonic skyrmion $(S_z=1)$ acquires phase factor $2\pi$ after
$2\pi$ rotation in original space. However, in \tat{} space we
have phase factor $\pi$ for each sector $l$, such that
$2\pi=\pi+\pi$. Together with the fact that spinor $(S_z={1}/{2})$
acquires phase factor $\pi$ after $2\pi$ rotation, this means that
the skyrmion \eqref{skyrmion_tilde} in \tat{} space is a fermion.

\section{BF theory}

In this section, after briefly reviewing 3D BF theory, we turn to
the study of the following possibility that a system in \tat{}
space is described by a 3D BF theory with $\tilde{k}=2$. We will
show that transformation of electromagnetic field operators
defined in section \ref{tilde_electrodynamics} is consistent with
quantization of charge in \tat{} space in $\sqrt{2}$ units of
elementary charge. This in turn means that a $\theta=2\pi$ axion
electrodynamics becomes a $\tilde{\theta}=\pi$ axion
electrodynamics in \tat{} space. Transformation of this model to
the original space gives a BF theory with $k=4$ and we argue that
this is an effective field theory for topological crystalline
insulators with $C_4$ symmetry.

\subsection{Review of 3D BF theory}

3D BF theory contains two statistical fields $a_\mu$ and
$b_{\mu\nu}$ coupled to quasiparticle and vortex currents
$$
    \mathcal{L}_{BF}=\frac{k}{4\pi}\varepsilon^{\mu\nu\rho\lambda}a_\mu
\partial_\nu b_{\rho\lambda}+j^\mu
a_\mu+\frac{1}{2}\Sigma^{\mu\nu}b_{\mu\nu}
$$
The first term imposes non-trivial statistics between
quasiparticles and vortices: when a quasiparticle $j$ is rotated
around a vortex $\Sigma$ the full wavefunction of the system
changes by a phase $\frac{2\pi}{k}$.\cite{bergeron} In the absence
of quasiparticles and vortices lagrangian can be written as
$$
    L_{BF}=\frac{k}{2\pi}\int d^3x a_i\dot{b}_i+a_0(\partial_ib_i)+b_{0i}(\varepsilon_{ijk}\partial_ja_k)
$$
where $b_i=\frac{1}{2}\varepsilon_{ijk}b_{jk}$. Integrating out
$a_0$ and $b_{0i}$ gives the constraints $\partial_ib_i=0$ and
$\varepsilon_{ijk}\partial_ja_k=0$ which can be resolved on a
torus by taking \cite{bfsc}
$$
    a_i=\frac{\bar{a}_i}{L_i}+\partial_i\Lambda,   \qquad
    b_i=\frac{\bar{b}_i}{S_i}+\varepsilon_{ijk}\partial_j\zeta_k
$$
where $\Lambda$ and $\zeta$ are periodic functions, $L_iS_i$ is
the volume of the system. The final lagrangian contains only
$\bar{a}_i$ and $\bar{b}_i$
$$
    L_{BF}=\frac{k}{2\pi}\bar{a}_i\partial_t{\bar{b}}_i
$$

It is possible to construct gauge invariant quantities from
$\bar{a}_i$ and $\bar{b}_i$, Wilson loops and Wilson
surfaces\cite{blau}
$$
    \mathcal{A}_i=e^{ie\bar{a}_i}, \qquad \mathcal{B}_i=e^{i\phi_0\bar{b}_i}
$$
This means that $\bar{a}_i\equiv \bar{a}_i+\frac{2\pi}{e}$ and
$\bar{b}_i\equiv \bar{b}_i+\frac{2\pi}{\phi_0}$ which is a
manifestation of quasiparticle and vortex number quantization.

\subsection{Effective field theory description of topological crystalline insulators}

$a$ and $A$ appear in the BF theory in the combination $a+A$, so
$a$ transforms in the same way as the electromagnetic potential.
However, $\bar{a}$ on a compact surface should be treated
differently because it is not possible to define $\bar{\tilde{a}}$
as the $K\rightarrow 0$ limit of the continuous expression.
Instead we write
\begin{equation}\label{bf_transformation1}
    \frac{k}{2\pi}\bar{a}_i\partial_t{\bar{b}}_i=\frac{\tilde{k}}{2\pi}\sum_{l}\bar{\tilde{a}}_i^{(l)}\partial_t{\bar{\tilde{b}}}_i^{(l)},
    \qquad \tilde{k}=\frac{k}{2}
\end{equation}
where
$\bar{\tilde{a}}_i^{(1)}=\bar{\tilde{a}}_i^{(2)}={\bar{a}_i}/{\gamma}$,
$\bar{\tilde{b}}_i^{(1)}=\bar{\tilde{b}}_i^{(2)}=\gamma\bar{b}_i$
for some numeric constant $\gamma$. Since Wilson loops and Wilson
surfaces in \tat{} space are given by
$$
    \tilde{\mathcal{A}}_i=e^{i\tilde{e}\bar{\tilde{a}}_i}, \qquad
    \tilde{\mathcal{B}}_i=e^{i\tilde{\phi}_0\bar{\tilde{b}}_i},
    \qquad \tilde{e}\tilde{\phi}_0=e\phi_0
$$
it follows that $\tilde{e}=\gamma e$. Thus
$\bar{\tilde{b}}_i=\gamma^2\cdot{\bar{b}_i}/{\gamma}$, which means
that transformation formulas for $\tilde{b}$ are given by that of
$\tilde{a}$ with an extra factor of $\gamma^2$, such that
\begin{equation}\label{bf_transformation2}
    \gamma^2\frac{\tilde{k}}{2\pi}\int_M{a}_i\partial_t{{b}}_i=
    \frac{\tilde{k}}{2\pi}\sum_{l}\int_{\tilde{M}}{\tilde{a}}_i^{(l)}\partial_t{{\tilde{b}}}_i^{(l)}
\end{equation}
For the two Eqs.\eqref{bf_transformation1} and
\eqref{bf_transformation2} to be consistent with each other
$\gamma$ must be equal to $\sqrt{2}$. This means that
$\tilde{e}=\sqrt{2}e$. The difference between transformation of
${a}$ and ${b}$ can be viewed as the consequence of charge-flux
duality, since transformation formulas for
$\tilde{a}$~($\tilde{b}$) differ from that of momentum $K$ by
additional factor $1/\sqrt{2}$~($\sqrt{2}$).

The full Lagrangian in \tat{} space together with time-reversal
symmetry breaking surface terms is given by the expression
$(\tilde{k}=2)$
\begin{equation}\label{tilde_lagrangian}
    \sum_{l}\int_{\tilde{M}}\frac{1}{2\pi}\varepsilon^{\mu\nu\rho\lambda}{\tilde{a}}_\mu^{(l)}\tilde{\partial}_\nu{{\tilde{b}}}_{\rho\lambda}^{(l)}+
    \frac{1}{2\pi}\varepsilon^{\mu\nu\rho\lambda}{\tilde{A}}_\mu^{(l)}\tilde{\partial}_\nu{{\tilde{b}}}_{\rho\lambda}^{(l)}+
    \frac{\tilde{e}^2}{8\pi}\varepsilon^{\mu\nu\rho\lambda}\tilde{\partial}_\mu{\tilde{a}}_\nu^{(l)}\tilde{\partial}_\rho{{\tilde{A}}}_{\lambda}^{(l)}
\end{equation}
Lagrangian \eqref{tilde_lagrangian} is two copies of the model
that was proposed in the Ref. \onlinecite{cho} as an effective
field theory for 3D topological insulators. Transformation of the
model \eqref{tilde_lagrangian} to the original space gives a model
with the Lagrangian density $(k=4)$
\begin{equation}\label{original_lagrangian}
    {\mathcal{L}}_{TCI}=\frac{1}{\pi}\varepsilon^{\mu\nu\rho\lambda}a_\mu
\partial_\nu b_{\rho\lambda}+\frac{1}{\pi}\varepsilon^{\mu\nu\rho\lambda}A_\mu
\partial_\nu b_{\rho\lambda}+\frac{e^2}{4\pi}\varepsilon^{\mu\nu\rho\lambda}\partial_\mu a_\nu
\partial_\rho A_{\lambda}
\end{equation}
An interesting consequence of Eq.\eqref{original_lagrangian} is
that vortex excitations in this model carry $\pi/2$-flux, i.e.
quarter flux quantum. This can be associated with $C_4$ symmetry,
since due to $C_4$ symmetry of the microscopic model it is
expected that quarter flux quantum must play a special role in the
theory.

Transformation of $\tilde{A}$, $\tilde{a}$ and $\tilde{b}$ under
time-reversal is the same as in original space:
$(\tilde{A}_0,-\tilde{A}_i)$, $(\tilde{a}_0,-\tilde{a}_i)$ and
$(-\tilde{b}_0,\tilde{b}_i)$ after time reversal, so the action
\eqref{original_lagrangian} is time-reversal invariant. $C_4$
rotation in original space interchanges the sectors with $l=1$ and
$l=2$ with each other.

If one integrates out $a$ and $b$ from Eq.
\eqref{original_lagrangian} then the remaining term is a $2\pi$
axion term. Eq. \eqref{axion_transformation}, together with the
fact that unit charge in \tat{} space is $\tilde{e}=\sqrt{2}e$,
gives that
$$
    \frac{e^2\theta}{2\pi h}\int_M
    \mathbf{E}\mathbf{H}=\frac{\tilde{e}^2 \tilde{\theta}}{2\pi
    h}\sum_{l}\int_{\tilde{M}}\tilde{\mathbf{E}}\tilde{\mathbf{H}},
    \qquad \tilde{\theta}=\frac{\theta}{2}
$$
Thus a $\theta=2\pi$ axion term transforms into two copies of
$\tilde{\theta}=\pi$ axion terms. These two copies are different,
because they involve Fourier components of electromagnetic field
with different momenta. So one can not combine these two BF
theories into a single BF theory with $\tilde{\theta}=2\pi$. These
two theories should be treated as independent unless there are
other terms mixing them with each other.

$Z_2$ classification of topological crystalline insulators can be
explained as it is done for 3D topological insulators, by applying
the arguments of Ref.\onlinecite{cho} to the model
\eqref{tilde_lagrangian}. Situation described here is, in some
sense, similar to the calculation of $Z_2$ invariants of
topological insulators, where Brillouin zone is divided into two
halfs, related to each other by time-reversal symmetry.\cite{roy2}
The difference is that here we deal with an effective field theory
instead of microscopic model: by a suitable transformation,
topological field theory is divided into two and a logic, familiar
from the study of 3D topological insulators,\cite{cho} is applied
to each half.

Eq. \eqref{original_lagrangian} is consistent with general
consideration of effective field theories of topological
insulators.\cite{dayi} It was shown in \onlinecite{dayi} that
before specifying the properties of space-time, the lagrangian
density of topological insulator is of BF type with two unknown
coefficients
$$
\Lambda_F(\varepsilon^{\mu\nu\rho\lambda}a_\mu
\partial_\nu b_{\rho\lambda}+\varepsilon^{\mu\nu\rho\lambda}A_\mu
\partial_\nu b_{\rho\lambda})+C_F\varepsilon^{\mu\nu\rho\lambda}\partial_\mu a_\nu
\partial_\rho A_{\lambda}
$$
This theory assumes that collective excitations in the bulk of a
topological insulator is given by neutral fermions, which couple
to external fields by electric and magnetic dipole moments. It is
reasonable to assume that collective excitations in topological
crystalline insulators also can be described by this method.

\section{Discussion}

We have presented some arguments that the theory
\eqref{original_lagrangian} might be an effective field theory for
recently proposed topological crystalline insulators \cite{fu} in
the case of $C_4$ symmetry. However the discussion has not been
rigorous and there are some points that should be cleared. For
example, transformation of momentum operators \eqref{momentum} was
defined only in the continuous limit and we do not know what are
the corresponding relations for a compact manifold. Such relations
are necessary for a rigorous theory. We also did not explain why
the model in \tat{} space is given by two copies of the model that
was proposed as a topological field theory for time-reversal
invariant topological insulators.\cite{cho} One of the reasons for
this choice was that this theory can easily explain $Z_2$ property
and we did not know any other theories. Though it was not a priori
stated that the theory must contain excitations with quarter flux
quantum, we obtained that there are such excitations making quite
reasonable propositions about the structure such a theory should
possess. This last circumstance makes these propositions more
plausible.

The other question is related to gapless surface excitations. It
was shown in the Ref.\onlinecite{cho} how one can account for
gapless surface excitations of 3D topological insulators in the
context of BF theory. Electronic degrees of freedom on the surface
are constructed using 2+1D bosonization scheme proposed in the
Ref.\onlinecite{aratyn,luther}. In principle, some analogous
procedure can be done in the case of fermions with gapless
quadratic spectrum in the continuous limit, by reducing them first
to fermions with linear spectrum. However it is more proper to
study fermions with quadratic spectrum from the beginning in view
of the following circumstance. In 2+1D in the continuum limit,
internal energy of a single two-component fermion with gapless
quadratic excitations is equal to internal energy of a boson with
gapless quadratic excitations due to the identity
$$
    2\int\frac{k^2d^2k}{e^{\alpha k^2}+1}=\int\frac{k^2d^2k}{e^{\alpha k^2}-1}
$$
analogous to equivalence of internal energies of bosons and
fermions in 1+1 D.\cite{haldane} No such simple relation exists
for 2+1D fermions and bosons with linear spectrum and bosonization
of Dirac fermions with linear spectrum in 2+1D, as it was
described in \cite{aratyn,luther}, is not fundamental.\cite{ls}

{\it Acknowledgements.} The author would like to acknowledge S.M.
Apenko and V.V. Losyakov for many useful discussions.




\begin{thebibliography}{5}

\bibitem{fu} L. Fu, Phys. Rev. Lett. 106, 106802 (2011).

\bibitem{fkm} L. Fu, C.L. Kane, and E.J. Mele, Phys. Rev. Lett. 98,
106803 (2007).

\bibitem{moore} J.E. Moore and L. Balents, Phys. Rev. B 75, R121306 (2007).

\bibitem{roy} R. Roy, Phys. Rev. B 79, 195322 (2009).

\bibitem{tft} X.-L. Qi, T. Hughes, and S.-C. Zhang, Phys. Rev. B 78,
195424 (2008).

\bibitem{emv} A.M. Essin, J.E. Moore, and D. Vanderbilt, Phys. Rev. Lett. 102, 146805 (2009).

\bibitem{ryu} P. Hosur, S. Ryu, and A. Vishwanath, Phys. Rev. B, 81, 045120
(2010).

\bibitem{cho} G.Y. Cho and J.E. Moore, Annals of Physics 326, 1515
(2011).

\bibitem{wilczek} F. Wilczek, Phys. Rev. Lett. 58, 1799 (1987).

\bibitem{vazifeh} M.M. Vazifeh and M. Franz, Phys. Rev. B 82, 233103 (2010).

\bibitem{qed} V.B. Berestetskii, E. M. Lifshitz, and L. P.
Pitaevskii, Quantum electrodynamics. Course of Theoretical
Physics. Vol. 4 (2nd ed.), London: Pergamon (1982).

\bibitem{goldstone} J. Goldstone and F. Wilczek, Phy. Rev. Lett. 47, 986
(1981).

\bibitem{chen} Y.H. Chen and F. Wilczek, Int. J. Mod. Phys. B 3, 117 (1989).

\bibitem{abanov} A.G. Abanov and P.B. Wiegmann, Phys. Rev. Lett. 86,
1319 (2001).

\bibitem{witten} E. Witten, Nucl. Phys. B 223, 433 (1983).

\bibitem{zee} F. Wilczek and A. Zee, Phys. Rev. Lett. 51,
2250 (1983).

\bibitem{wu} Y.S. Wu and A. Zee, Phys. Lett. B 147, 325 (1984).

\bibitem{moon} E.G. Moon, arXiv:1202.5389.

\bibitem{bergeron} M. Bergeron, G.W. Semenoff, and R.J. Szabo, Nucl. Phys. B 437, 695 (1995).

\bibitem{bfsc} T.H. Hansson, V. Oganesyan, and S.L. Sondhi, Annals of Physics 313, 497
(2004).

\bibitem{blau} M. Blau and G. Thompson, Ann. Phys. 205, 130 (1991).

\bibitem{roy2} R. Roy, Phys. Rev. B 79, 195321 (2009).

\bibitem{dayi} O.F. Dayi and M. Elbistan, arXiv:1205.1967v1.

\bibitem{aratyn} H. Aratyn, Phys. Rev. D 28, 2016 (1983); Nucl. Phys. B 227, 172 (1983).

\bibitem{luther} A. Luther, Phys. Rev. B 19, 320 (1979); Phys. Rep. 49, 261 (1979).

\bibitem{haldane} F.D.M. Haldane, J. Phys. C 14, 2585 (1981).

\bibitem{ls} A. Luther and K.D. Schotte, Nucl. Phys. B 242, 407
(1984).

\end{thebibliography}
\end{document}